\documentclass[achemso,amsmath,amssymb,reprint,]{revtex4-1}
\usepackage{graphicx}
\usepackage{dcolumn}
\usepackage{bm}
\usepackage[utf8]{inputenc}
\usepackage{natbib}
\usepackage[T1]{fontenc}
\usepackage{mathptmx}
\usepackage{etoolbox}

\makeatletter 
\def\@email#1#2{%
 \endgroup
 \patchcmd{\titleblock@produce}
  {\frontmatter@RRAPformat}
  {\frontmatter@RRAPformat{\produce@RRAP{*#1\href{mailto:#2}{#2}}}\frontmatter@RRAPformat}
  {}{}
}%
\makeatother
\usepackage{hyperref}
\hypersetup{colorlinks,linkcolor={blue},citecolor={blue},urlcolor={blue}}
\begin{document}

\title {Achievable Information-Energy Exchange in a Brownian Information Engine through Potential Profiling}

\author{Rafna Rafeek}
\affiliation{Department of Chemistry and Center for Molecular and Optical Sciences \& Technologies, Indian Institute of Technology Tirupati, Yerpedu 517619, Andhra Pradesh, India}
\author{Debasish Mondal}
  \email{debasish@iittp.ac.in}
\affiliation{Department of Chemistry and Center for Molecular and Optical Sciences \& Technologies, Indian Institute of Technology Tirupati, Yerpedu 517619, Andhra Pradesh, India}

\date{\today}
\begin{abstract}
The information engine extracts work from a single heat bath using mutual information obtained during the operation cycle. This study investigates the influence of the potential shaping in a Brownian information engine (BIE) in harnessing the information from thermal fluctuations. We have designed a BIE by considering an overdamped Brownian particle inside a confined potential and introducing an appropriate symmetric feedback cycle. We find that the upper bound of the extractable work for a BIE with a monostable centrosymmetric confining potential, with a stable state at the potential centre, depends on the bath temperature and the convexity of the confinement. A concave confinement is more efficient for an information-energy exchange. For a bistable confinement with an unstable centre and two symmetric stable basins, one can find an engine-to-refrigeration transition beyond a certain barrier height related to the energy difference between the energy barrier and the stable basins. Finally, we use the concavity-induced gain in information harnessing to device a BIE in the presence of a multistable potential that can harvest even more energy than monostable confinement. 
\end{abstract}
\maketitle
\noindent \emph{Introduction:} Living systems are inherently subjected to fluctuations that can be thermal, chemical, or others in nature, and these fluctuations are rectified to achieve several fascinating functional biophysical outcomes \cite{fang2019}. During a state change, fluctuations lead to changes in the surprisal (information) of the event. These fluctuations, which underpin biological activity, lead to an exchange of information and energy with the environment. For example, biological mesoscopic devices harness information about fluctuations to extract energy from the noisy environment \cite{Heiner_2021, borsley_2021,chakraborty2023}. The link becomes a fundamental challenge in nonequilibrium statistical physics since the celebrated thought experiment of "Maxwell's demon" was introduced \cite{Szilard1929zphys, Landaueribm1961, rex2017maxwell}. This demon inspects gas molecules in a single heat bath and utilizes the obtained information to extract work, thus emanating the notion of the information engine. The concept thus sparked a new interest in exploring the link between information and thermodynamics.

In light of the recent development of stochastic thermodynamics \cite{Jarzynski1997prl, sekimoto2010stochastic, seifert2008}, an information engine can be considered a practical realization of Maxwell's demon of a mesoscopic scale. Nowadays, the information engine can be devised as a feedback control that extorts work from a single heat bath using the mutual information earned during the measurement \cite{Sagawa2010prl, Toyabe2010natphys, Lopez2008prl, Paneru2018prl,  paneru2020, Ashida2014pre, Pal2014pre,  Bauer_2012, saha2021, Paneru_2022, malgaretti2022, ali2022geometric, Abreu2011epl,  Mandal_2013, Taichi_2013, rafeek2023geometric, Paneru2018prl, Park2016pre, Berut2012nat,paneru2021transport}. During a state change of such a small-scale information engine, where the thermal fluctuations are significant, the average work done and the related information change are found to be inter-related as:  $ -\langle W \rangle \leq -\Delta F + k_BT( \langle I \rangle - \langle I_{u} \rangle)$ \cite{Sagawa2010prl, Ashida2014pre}.
 Where  $T$ is the temperature (fluctuation strength), $k_B$ is the Boltzmann constant and $\langle .. \rangle$ signifies the notation of the ensemble averaging. We consider the work done by the system as a negative quantity. The $\langle I \rangle$ and $\langle I_u \rangle$ denote the information gained during the measurement and the loss of information due to the post-feedback relaxation process, respectively. If $P (Y)$ is the probability of a random event $Y$, the information (surprisal) related to the process is defined as: $I=-lnP(Y)$. The inequality thus captures the exchangeability of information and energy during a state change and shows that the average extraction of the work $(-\langle W \rangle)$ cannot exceed the sum of the free energy differences $(-\langle \Delta F \rangle)$ and the available information $(\langle I\rangle - \langle I_u \rangle)$. We will introduce the definition and significance of $\langle I\rangle$ and $\langle I_u \rangle$ in detail later. 
\begin{figure*}[!htb]
\begin{minipage}[b]{0.29\linewidth}
\centering
\includegraphics[width=1\textwidth]{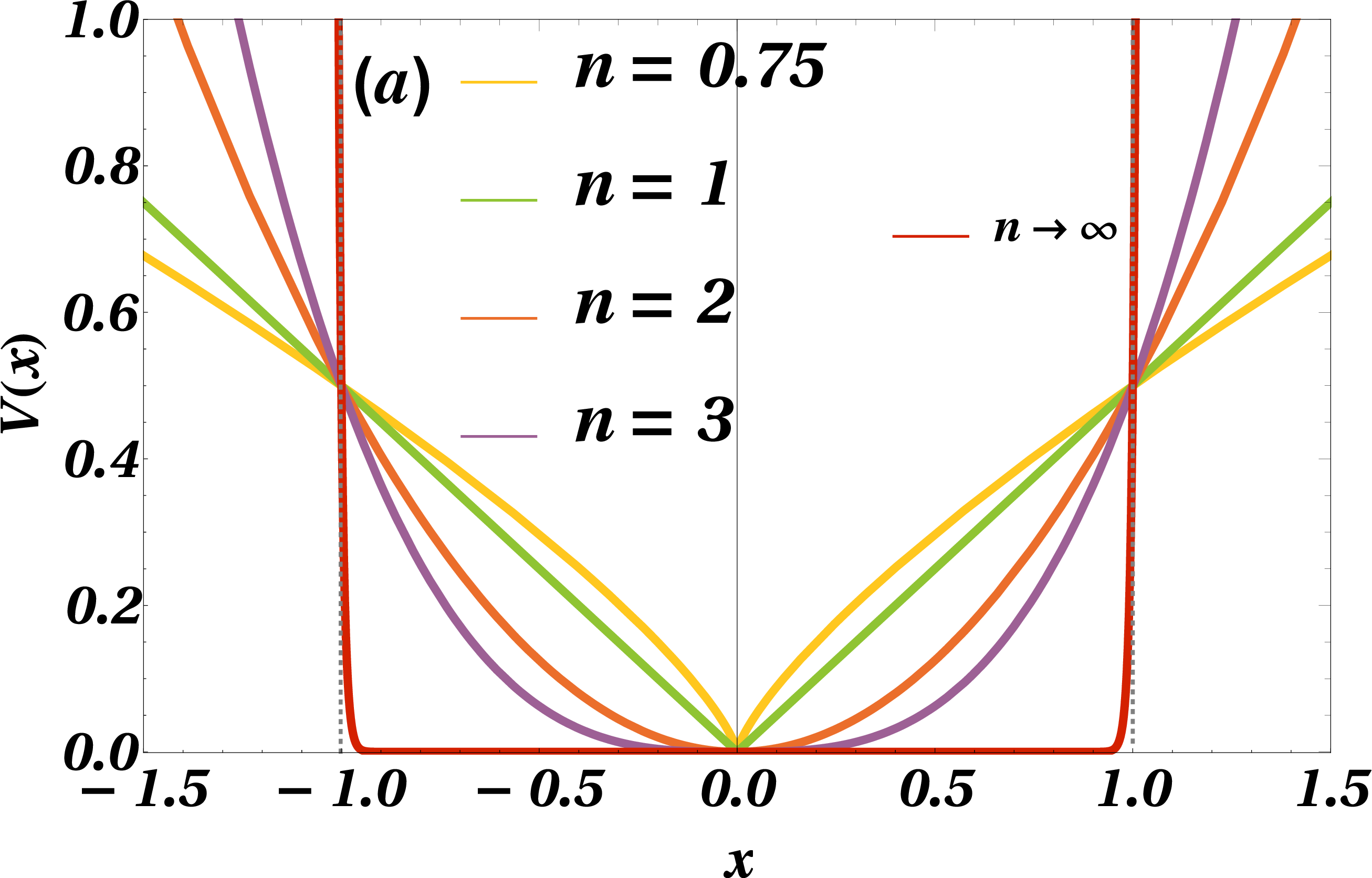}
\end{minipage}
\hspace{0.01cm}
\begin{minipage}[b]{0.29\linewidth}
\centering
\includegraphics[width=1\textwidth]{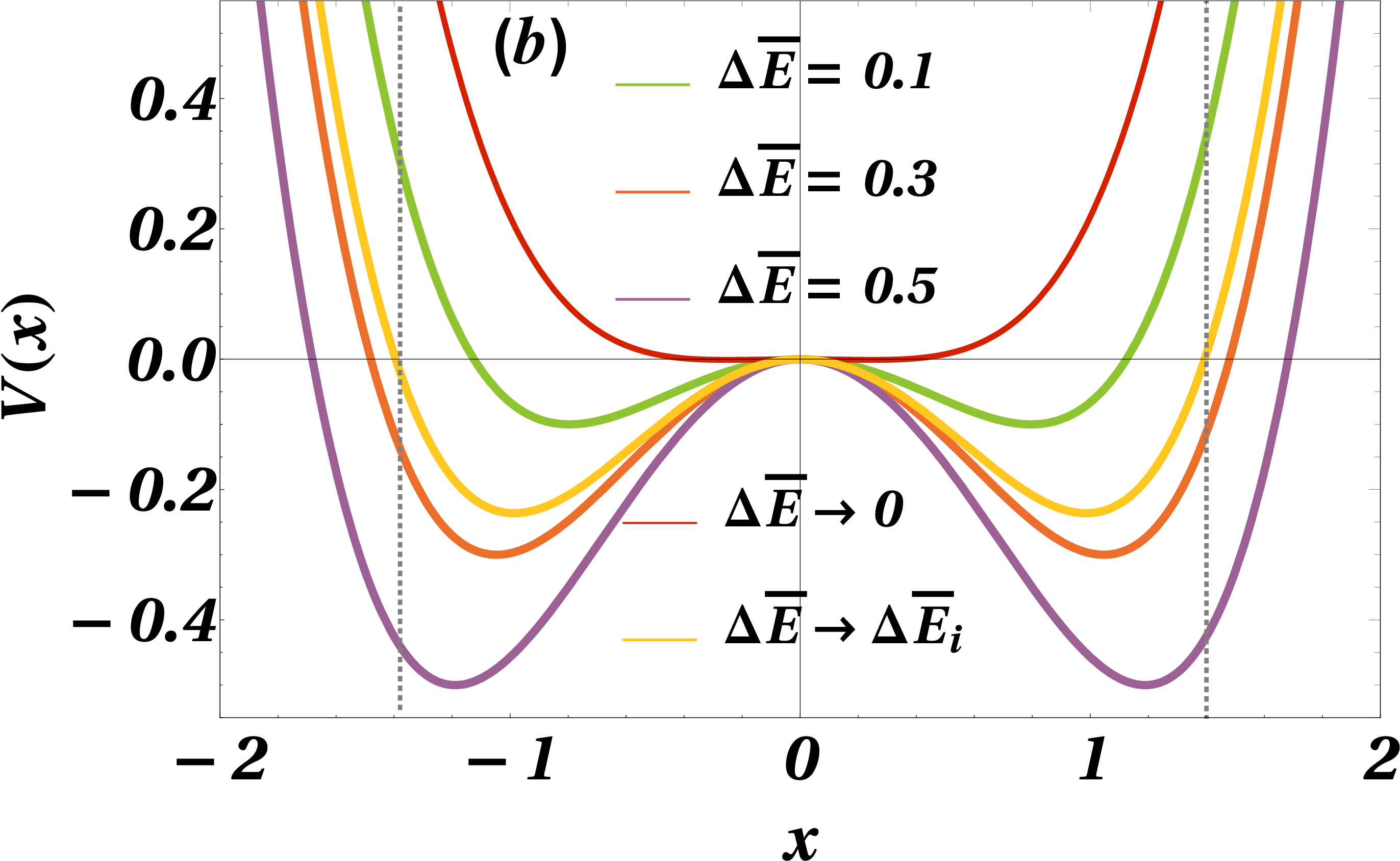}
\end{minipage}
\hspace{0.01cm}
\begin{minipage}[b]{0.29\linewidth}
\centering
\includegraphics[width=0.97\textwidth]{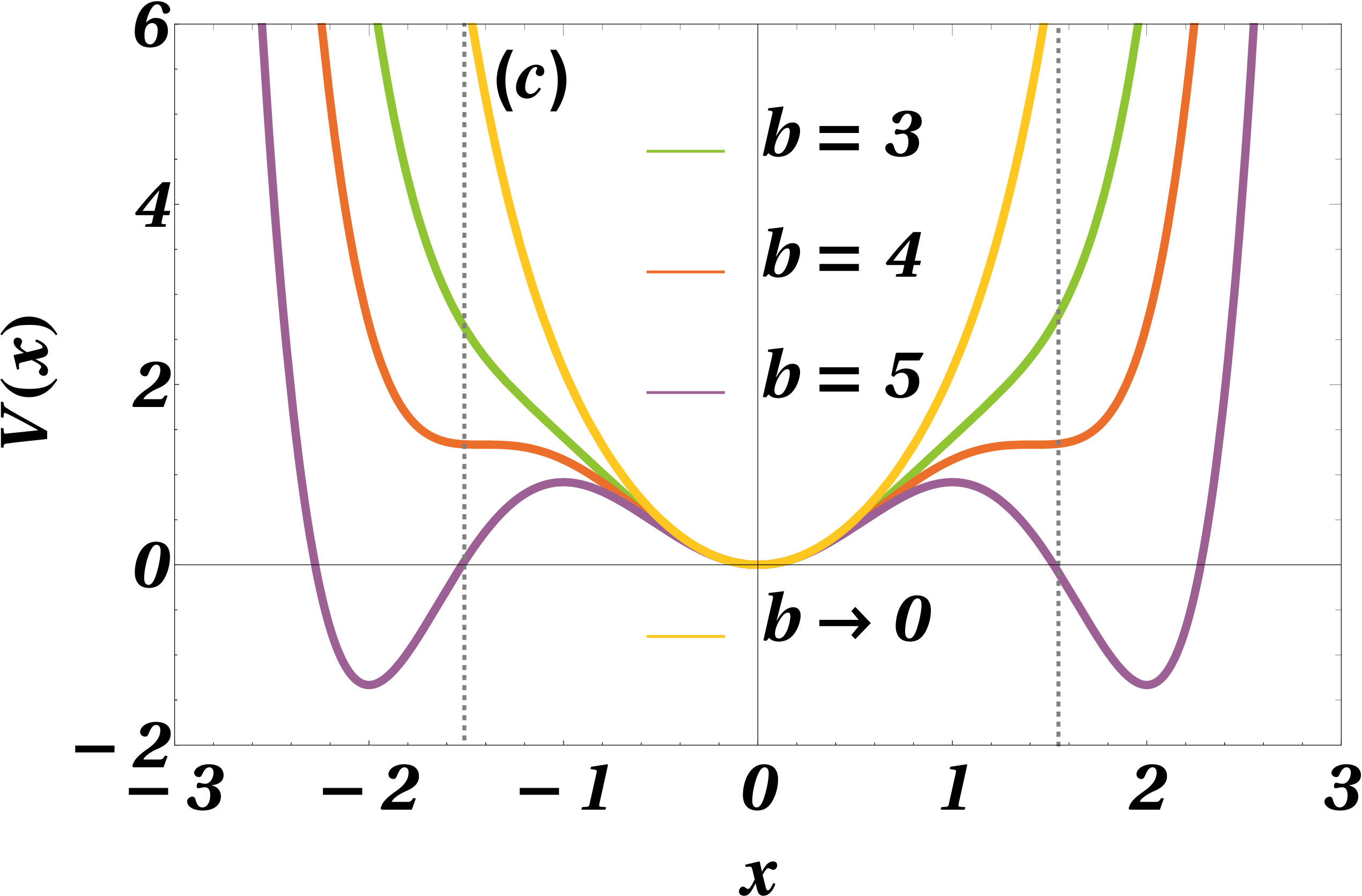}
\end{minipage}
\caption{(a) Monostable potential  $(V(x)=a|x|^n)$ shaping for different values of exponent ($n$), and with $a=0.5$.
(b) The bi-stable potential $(V(x)= -\frac{a}{2}x^2+\frac{b}{4}x^4)$ variation with different values of scaled barrier height $\Delta \overline{E}$ ($\Delta E = a^2/4b$), with $b=1$.
(c) The tripple well potential of the form $V(x)= \frac{a}{2}x^2-\frac{b}{4}x^4 + \frac{c}{6}x^6$ for different different values of $b$, and with $a=4$ and $c=1$. }
\label{f1}
\end{figure*}
 
Persuaded by an information-energy exchange in small-scale living systems, numerous theoretical models of stochastic information engines operating within a single thermal reservoir, from classical \cite{Abreu2011epl, Bauer_2012, Mandal_2013, Taichi_2013, Pal2014pre, Ashida2014pre, ali2022geometric, rafeek2023geometric} to quantum \cite{ Kim2011prl, Bruschi2015pre, Goold2016jphysA, koski2014pnas, malgaretti2022}, have been explored and validated through experiments \cite{Toyabe2010natphys, Berut2012nat, Park2016pre, paneru2021transport, Paneru2018prl, paneru2020, Lopez2008prl}. An overwhelming majority of these studies examine Brownian information engines composed of overdamped particles trapped in the external harmonic potential $(V(x)=\frac{1}{2}kx^2)$ (where $k$ is the stiffness of the potential), undergoing feedback control under the influence of a thermal bath \cite{Pal2014pre, Ashida2014pre, Park2016pre, Paneru2018prl, paneru2020}.  The potential profiles in these situations act as monostable centrosymmetric confinement with a stable state at the centre ($x=0$) \cite{Abreu2011epl, Bauer_2012, Mandal_2013, Taichi_2013, Pal2014pre, Ashida2014pre, ali2022geometric, rafeek2023geometric, Park2016pre, Paneru2018prl, Lopez2008prl, paneru2020}. One of the common ways to introduce a symmetric feedback control to get the energy-information exchange relation is as follows. Once the system reaches an equilibrium state, the particle's position is measured ($x'$), and the centre of the potential changes instantaneously to the measured location (from $x=0$ to $x=x'$ position). This aforementioned potential set-up and the feedback protocol give rise to the upper bound of achievable work of such an engine as the error-free work output equals the average potential energy of the harmonic trapping ($=\frac {1}{2} k_BT$), which is independent of potential stiffness \cite{Ashida2014pre}. Our recent study investigated a Brownian information engine without external confining potentials \cite{ali2022geometric, rafeek2023geometric} but subjected to a monostable entropic enclosure \cite{zwanzig1992, rubi2001, reguera2006, ali2024optimizing, mondal2010activation}.
 The upper bound of work extraction from such a geometric Brownian information engine shows a crossover from $\left(5/3 - 2 \ln 2\right)k_BT$ to $\frac{1}{2} k_BT$ while transitioning from an entropy-dominated regime to an energy-dominated one \cite{ali2022geometric}. In either situation, the potential energy difference is linked to the extractable work for a single trajectory during the feedback $(-W(x') = V(x') - V(0))$. Thus, the conversion of information into extractable work requires constraining the Brownian particle within an external potential (or effective entropic potential) and a stability gain in terms of the potential energy difference under the feedback control.
 
 The shape of the external confining potentials often plays a crucial role in energy harnessing from a stochastic environment (like Brownian motion). Studies show that noise-assisted barrier crossing rate \cite{marie2020} and different noise-induced phenomena like stochastic resonance  \cite{mcnamara1989, mondal2016resonance, ghosh2007, zhu2022}, resonant activation \cite{doering1992, reimann1995, jan1996, wagner1999, bart2002,mondal2010activation, fisconaro2011, chattoraj2014dynamics,mondal2012entropic} and ratchet rectification \cite{magnasco1993, zwanzig1988, mondal2009roughratchet, kato2013quantum, mondal2016ratchet} display advantageous and non-trivial behaviour when exposed to potential landscapes of particular shapes. Therefore, it is noteworthy to investigate the impact of different potential profiles on mutual information-energy exchange in fluctuating environments. In particular, several questions need to be addressed: a) Can the achievable upper bound ($\ \frac{1}{2} k_BT$ for a harmonic trapping) be affected by tuning the stiffness of the monostable centrosymmetric confinement? b) How does switching to a centrosymmetric potential with an unstable centre influence the information-energy exchangeability? c) Does introducing a centrosymmetric potential with multiple stable states impact information gain and loss during relaxation? d) Moreover, can we improve the engine performance beyond the efficiency of a monostable confining potential by shaping it into a multistable potential trapping? 

 To bridge this gap, we consider an overdamped Brownian particle confined within a single-well potential that adheres to a power-law form $ V(x)= a|x|^n$, where $a$ is a constant parameter, and the exponent $n$ can be any positive integer number. Here, $|Y|$ means the modulus (absolute value) of the variable $Y$. We employ an appropriate symmetric feedback protocol in the spirit of \cite{Ashida2014pre, ali2022geometric}, as explained in Supporting Information (SI). To comprehend the impact of potential with an unstable centre, we investigate a gradual transition from a monostable to a bistable potential. This transition can be facilitated by tuning the control parameter of a quartic potential that introduces an unstable centre and two equidistant iso-energetic wells. To further understand the influence of multi-stable trapping on the information-energy conversion, we examine a continuous regulation of a mono-stable to a tri-stable potential, always maintaining a stable state at the origin. It is noteworthy that the recent developments of experimental techniques enable us to mimic the aforementioned model potential of different shapes with the help of an automated microfluidic experiment that uses holographic optical tweezers \cite{Paneru2018prl, marie2020} under overdamped (not limited though) conditions.\\
 \begin{figure*}[!htb]
\begin{minipage}[b]{0.29\linewidth}
\centering
 \includegraphics[width=1\textwidth]{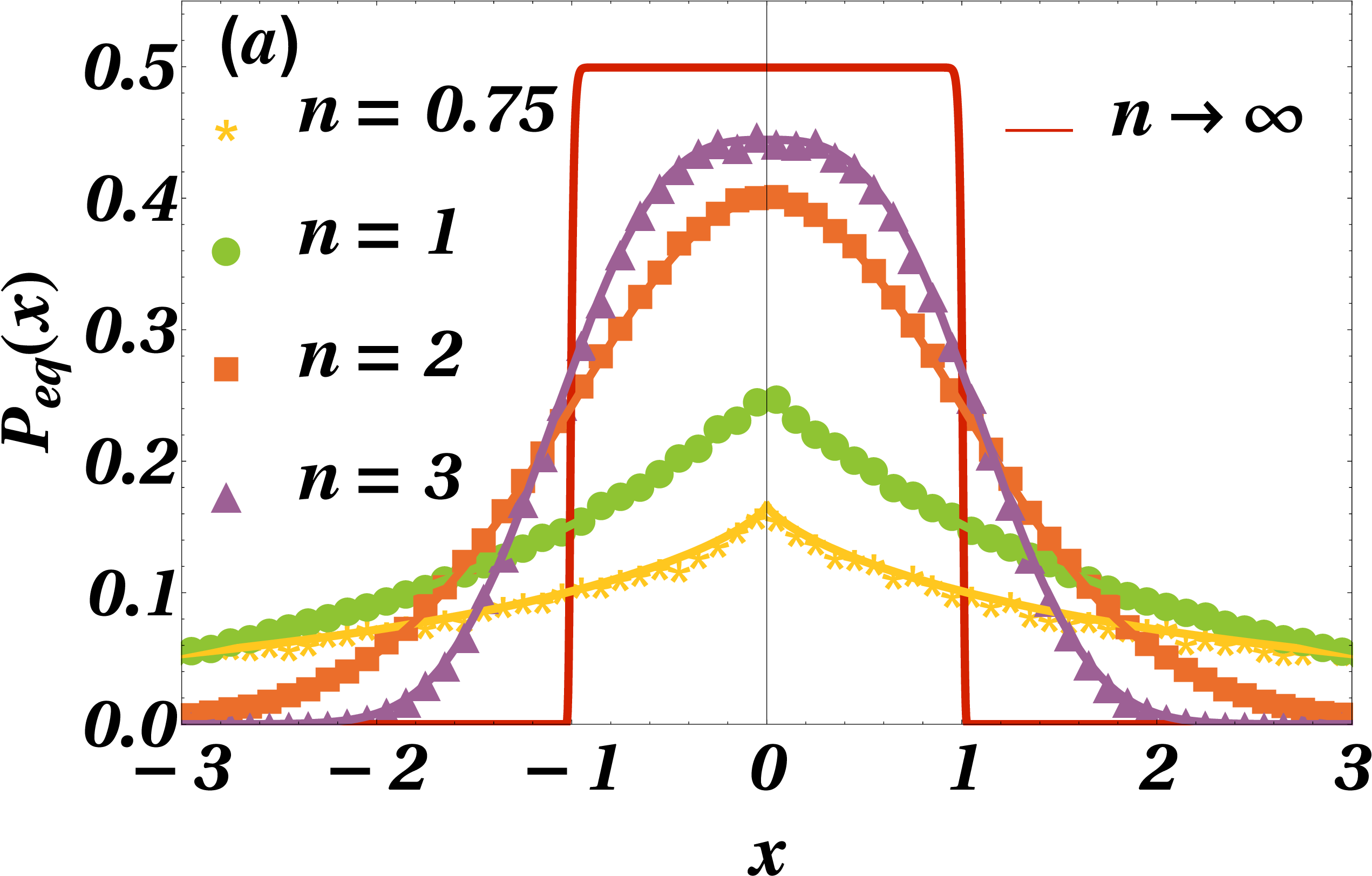}
\end{minipage}
\hspace{0.01cm}
\begin{minipage}[b]{0.29\linewidth}
\centering
\includegraphics[width=1\textwidth]{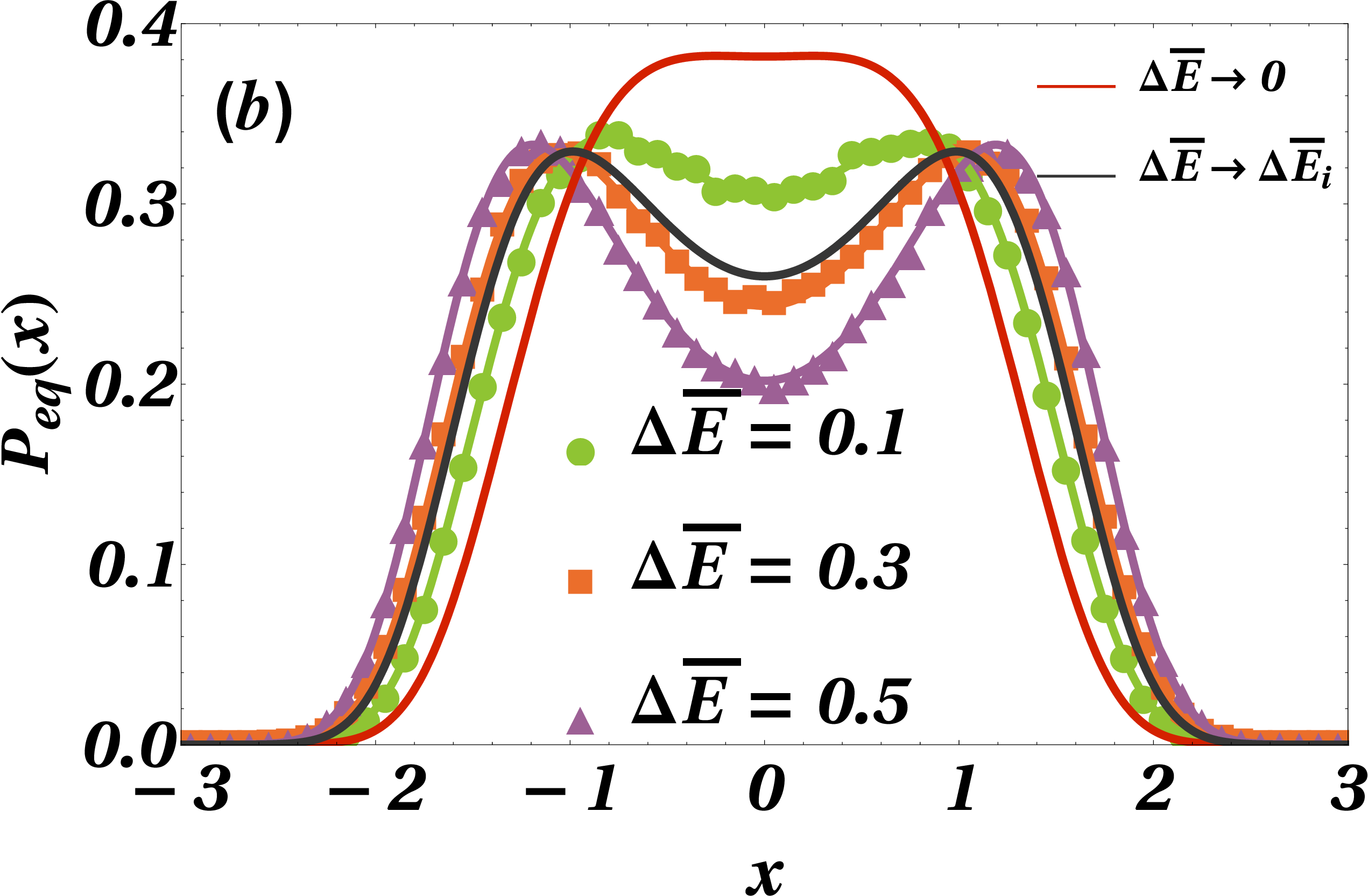}
\end{minipage}
\hspace{0.01cm}
\begin{minipage}[b]{0.29\linewidth}
\centering
\includegraphics[width=1\textwidth]{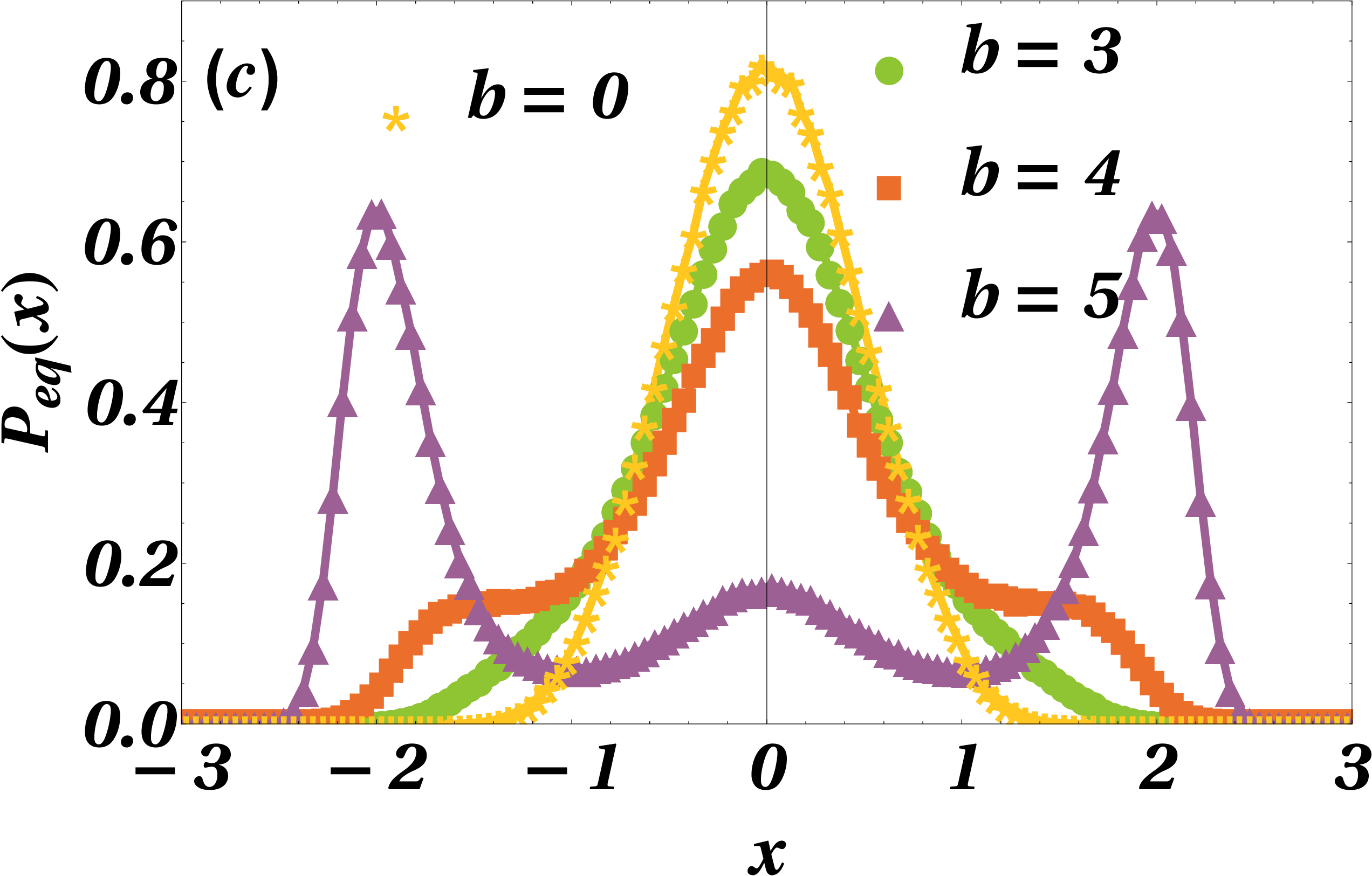}
\end{minipage}
\caption{The steady-state (equilibrium here) probability distribution ($P_{eq}(x)$) for a particle confined in (a) mono-stable potential with different values of $n$ for $a=0.5$. 
(b) Bistable potential with different values of scaled barrier height $\Delta \overline{E}$. 
(c) Tripple well potential with different values of $b$. In all cases, solid lines represent the theoretical predictions and points are obtained from numerical simulation (Eq.~\ref{1}). The potential parameters are taken same as Fig. \ref{f1}.}
\label{f2}
\end{figure*}

\noindent \emph{Modelling the Brownian Information Engine:} An overdamped Brownian particle, confined in an external centrosymmetric potential, $V(x-\lambda$), centred at $\lambda$ is under consideration that follows the following 1D  Langevin equation:
\begin{equation}\label{1}
\gamma \dot{x} = -V^{'}(x-\lambda) + \sqrt{2D} \gamma \eta(t), 
 \end{equation}
where $\gamma$ is the frictional coefficient, and $\eta(t)$ is Gaussian white noise, described as $\left \langle \eta (t)\right \rangle = 0$ and $\left \langle \eta(t) \eta ({t}')\right \rangle =\delta (t-{t}')$. The diffusion coefficient $(D)$ characterizes the noise strength $(D=k_BT / \gamma)$. The stationary state distribution of particle position (here equilibrium as well) reads: $P_{eq}(x) = \mathcal{N} \exp \bigg [ -\frac{V(x -\lambda ) }{k_B T} \bigg]$, where $\mathcal{N}$ is normalization constant. 
 The forward feedback protocol involves measurement, feedback, and relaxation steps \cite{supple}. At the start of the cycle ($t=0$), the system is in thermal equilibrium with the potential centred at \( \lambda = 0 \). 
 During the measurement step, the particle's position is measured precisely as $x(t) = x'$. In the feedback step, the potential centre is instantly shifted from $\lambda = 0$ to $ \lambda = x'$. The particle then relaxes to thermal equilibrium (with $\lambda = x'$), and the cycle repeats. We have assumed an error-free measurement, and the potential shift is instantaneous. Thus, the potential energy difference can be linked to work during the feedback, \(-W(x') = \Delta V(x' - \lambda) = V(x') - V(0)\). The average work obtained from the information engine is defined as:
 \begin{equation}\label{2}
 -\langle W \rangle = \int_{-\infty}^{\infty}dx' \Delta V(x'-\lambda) P_{eq}(x').
  \end{equation}
 Next, we evaluate the average information gathered during this feedback step as, $ \langle I \rangle =- \int_{-\infty}^{\infty}dx P_{eq}(x)\ln P_{eq}(x)$. During the relaxation process, a certain amount of gathered information is lost, however.  One can estimate this loss as the unavailable information, $ \langle I_u \rangle =- \int_{-\infty}^{\infty}dx  P_{eq}(x) \ln P_{eq}(0)$.
 To evaluate unavailable information, one needs to consider a backward protocol. Both the forward and backward protocol with a particle trapped inside a single-well confinement is illustrated in \cite{supple}. It is straightforward to show that the proposed feedback mechanism forms a lossless information engine when the feedback controllers operate in an error-free environment. The achievable work, thus can be bounded in terms of available information as: $-\langle W \rangle = k_BT ( \langle I \rangle -\langle I_u \rangle)$. One can also verify the integral fluctuation theorem by substituting the work and information terms in the following relation:
\begin{equation}\label{3}
\begin{aligned}
      \langle \text{e}^{-(\overline{W}+I-I_u )} \rangle =
    \int_{-\infty}^{\infty} {dx} P_{eq}({x})\text{e}^{\overline{V}(x)-\overline{V}(0)}\frac{P_{eq}({x})}{P_{eq}(0)}
     \;\;= 1.
\end{aligned}
\end{equation}
Here onwards, for any variable $Z$, we consider $\overline{Z}=\frac{Z}{k_B T}$ unless mentioned otherwise.

\begin{figure}[!htb]   {\includegraphics[width=0.43\textwidth]{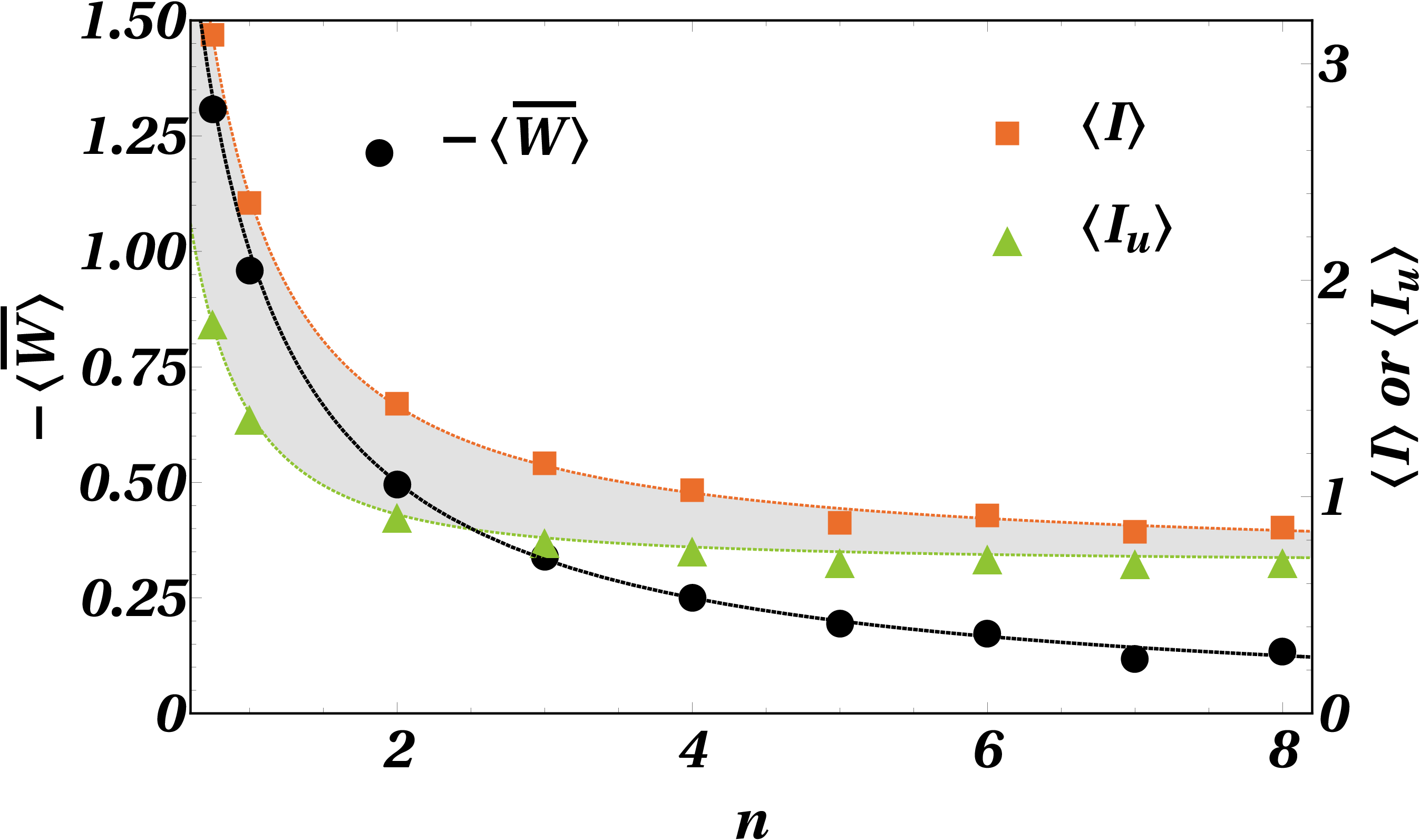}} 
     \caption{The extractable scaled work $(-\langle \overline{W} \rangle)$, information $(\langle I \rangle)$ and unavailable information $(\langle I_u \rangle)$ as a function of $n$. The parameter set is chosen: $a=1$, and $D=1$. The points obtained from numerical simulation (Eq.~\ref{1}) are in excellent agreement with theoretical predictions (lines).}
    \label{f3}
\end{figure}

\begin{figure}[!htb]
    {\includegraphics[width=0.43\textwidth]{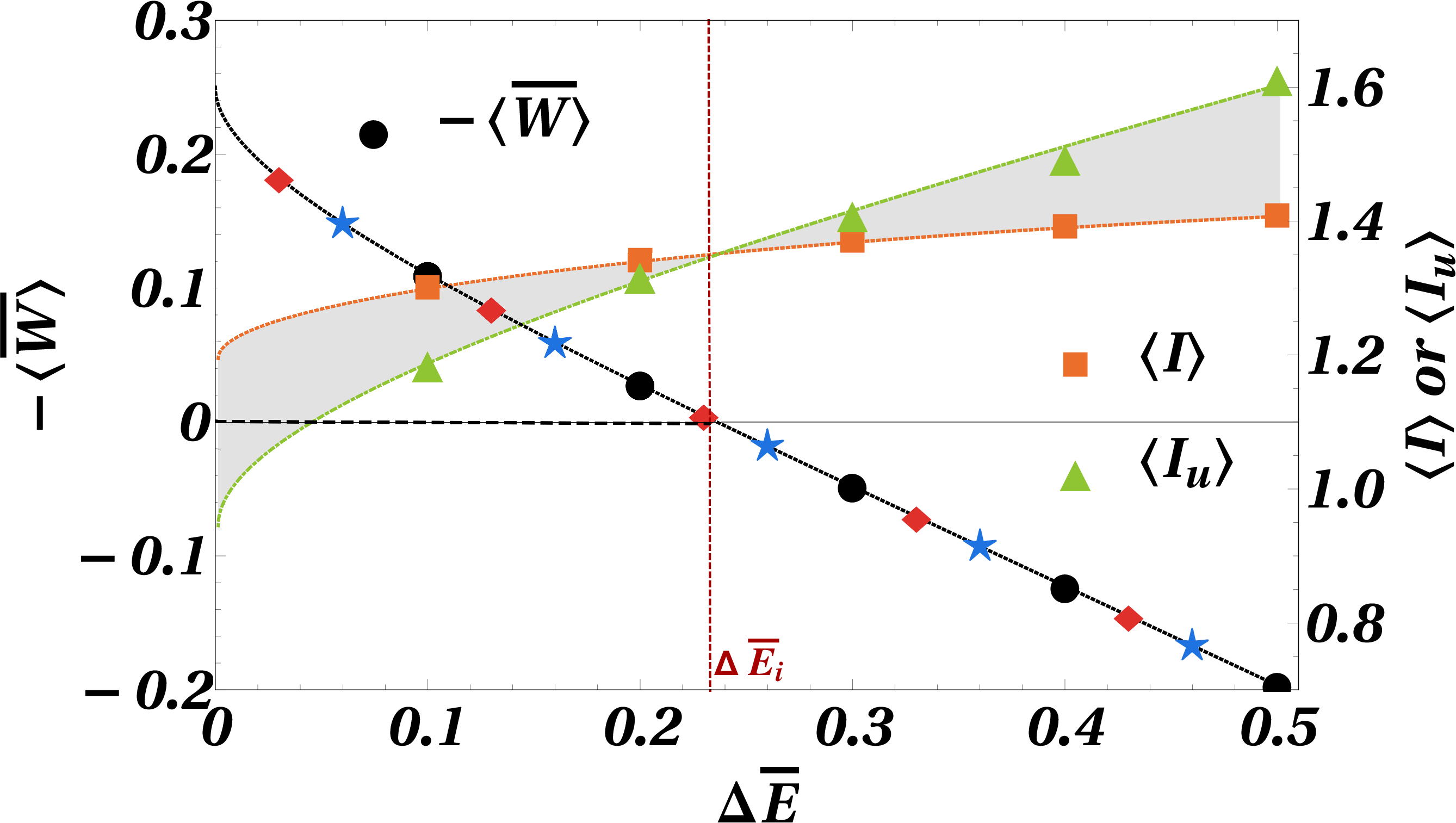}} 
     \caption{The extracted work $(-\langle W \rangle)$, information $(\langle I \rangle)$ and unavailable information $(\langle I_u \rangle)$ as function of $\Delta \overline{E}$. 
     The points obtained from numerical simulation (Eq.~\ref{1}) are in excellent agreement with the theoretical prediction. }
    \label{f4}
\end{figure}

We follow the experimental values (from studies \cite{Paneru_2022, paneru2021transport}) as a reference scale to derive the dimensionless description of system parameters. Henceforth, characteristic reference energy scale $(E_r)$ is equivalent to $k_B T_R$ at $T_R = 293 K$ and reference unit of frictional coefficient $(\gamma_r)$ set to be $18.8 \; nNm^{-1}s$. We scale all lengths by $(x_r) \sim 20 \; nm$ and time $(t_r)$ by the reference scale of $\sim 2 \;ms$. Consequently, the reference unit of stiffness for the harmonic potential is $E_r / x_r^{2}$, i.e. $10 \; pN \; \mu m^{-1}$. Similarly, the potential parameters can be derived for other landscapes. In this spirit, we scale the energy as $\tilde{E}=E/E_r$ and the frictional coefficient as $\tilde{\gamma}=\gamma / \gamma_r$. The length scaled as $\tilde{x} = x/x_r$ and time $\tilde{t} = t/t_r$. For clarity, we will omit the tilda symbol in the following sections. We numerically compute the work done during the feedback and the equilibrium probability distribution using the overdamped Langevin equation (Eq.~\ref{1}).
The particle's trajectory is simulated using an improved Euler's method \cite{hildebrand1987} with a time step of $10^{-3}$ units. The Gaussian white noise is introduced via the Box-Muller algorithm \cite{box1958}. Unless stated otherwise, we set the $\gamma =1$ and $k_B T=1$ and generate $\sim 10^{7}$ trajectories to calculate the average physical observables. Wherever an exact analytical integration is not doable, we perform a numerical integration by using Simpson's $1/3$rd rule \cite{hildebrand1987} to get the theoretical prediction. Other parameter details are provided either in the figure captions or in \cite{supple}. As mentioned earlier, we primarily focus into the influence of confining potential profiling on the extraction of work from the engine comprised of particles within three distinct situations: (a) inside monostable trapping where we tune the shape following a power law type potential, (b) continuous regulation of a from single-well trapping to a bistable one creating instability at the potential center, and (c) altering a single-well trap to a triple-well one, always maintaining a stable state at the potential center.\\
\noindent \textit{Profiling convexity (concavity) of a monostable confinement:} To begin with, we consider a Brownian particle confined in a monostable potential of the form $V(x)=a|x|^n$, where $a$ is the force constant and the exponent is always positive, $n > 0$. The shape of the potential is thus concave, linear, and convex for $n < 1$, $n=1$, and $n > 1$, respectively (Fig.~\ref{f1} (a)). The convexity (concavity) continuously increases with increasing (decreasing) $n$. The energy minimum always stays at the potential centre for any values of $n$.
 The stationary distribution (equilibrium here) of particle position reads as:
\begin{equation}\label{4}
    \begin{aligned}
      P_{eq}(x)= \mathcal{N}_m \exp  [- \overline{a}|x|^n  ], \; \text{with} \;
 \mathcal{N}_m =   \frac{\overline{a}^{\frac{1}{n}}}{2 \Gamma (1+\frac{1}{n})}.
    \end{aligned}
\end{equation}
Here, $\Gamma(z)$ denotes the gamma function of the form $\Gamma(z) = \int_0^\infty t^{z-1} e^{-t} dt$. The $P_{eq}(x)$ follows a generalized unimodal symmetric distribution and exhibits different degrees of peakedness and tail behavior with varying exponent $(n)$ (Fig.~\ref{f3}). For $n=2$, the $P_{eq}(x)$ is a standard normal distribution (corresponding to harmonic potential). For $n > 2$, the $P_{eq}(x)$ have shorter tails and a taller peak near the potential centre. When $n \to \infty$, the monostable trapping looks like a box potential ($V(x) \sim 0,|x|<1$ else, $V(x)=\infty$), and one could expect a rectangular distribution, i.e. finite probability for $|x|<1$ and zero otherwise. For $n \to 1$, the $P_{eq}(x)$ converges to the Laplace distribution (following the linear potential).  
 We then find the achievable extractable work using the distribution ($P_{eq} (x)$) obtained from Eq.~\ref{4} as: $-\langle \overline{W} \rangle = \int_{-\infty}^{\infty}dx' \Delta V(x'-\lambda) P_{eq}(x') = \frac{1}{n}$. Such dependence of $-\langle \overline{W} \rangle$ on the potential exponent can be explained by scrutinising the potential set-up and the feedback protocol carefully. It is because of the location of the confinement centre and the spatial symmetry of the feedback site, the upper bound of $-\langle \overline{W} \rangle$ is always set by the average potential energy of the monostable trapping.
 Thus, the work output is inversely proportional to the power exponent and extent of information energy conversion is more inside a concave potential ($n<1$) than that of a convex trapping ($n>1$). The variation of the average extracable work with a varying exponent is shown in Fig.~\ref{f3}. For a harmonic trapping $(n=2)$, thus $\frac{1}{2} k_BT$ amount of work $(-\langle \overline{W} \rangle = \frac{1}{2})$ is obtained as expected. Notably, for concave potential surface $(n<1)$ the work extraction exceeds $k_B T$. 
  Using the equilibrium probability distribution (Eq.~\ref{4}), one can also estimate the average acquired information and unavailable information. The net information acquired from the measurement can be derived as $\langle I \rangle = -\ln [\mathcal{N}_m ] + \frac{1}{n}$ and $\langle I_u \rangle = -\ln [\mathcal{N}_m ]$, respectively. As depicted in Fig~\ref{f3}, both the average acquired information $(\langle I \rangle)$ and unavailable information $(\langle I_u \rangle)$ decreases with increasing power exponent $(n)$. This can be attributed to the surprisal related to the particle position. With an increase in $n$, the convexity of the confinement increaes (concavity decreases). The particle diffuses further from the potential centre (heavy tail), thus raising overall positional uncertainty. Due to the increased uncertainty in position, a higher amount of information is gained in measurement and lost during the relaxation process. However, the available (net) information converted into extracted work $(\langle I \rangle - \langle I_u \rangle)$ is always $1/n$, showing that the relative increase in acquired information during the measurement step is more.\\ 
\noindent \textit{A bistable trapping with an unstable potential centre:} As one can notice from the previous outcomes, the location of the stablest point in the confinement is crucial for an efficient net information gain over a feedback cycle.
 Therefore, we consider a centro-symmetric bistable potential with an unstable centre for the next. One standard way to create such a profile is through a quartic potential of the form: $ V(x)= -\frac{a}{2}x^2 + \frac{b}{4} x^4$, where $a$ and $b$ are the structural parameters of potential. This potential has two minima (at $x=\pm \sqrt{\frac{a}{b}}$) separated by a maximum (at  $x=0$), creating a symmetric double-well shape with an energy barrier of $\Delta E = a^2/4b$. One can alter the value of $a$ from a zero to a non-zero to modulate the potential landscape from a mono-stable to a bi-stable shape (Fig.~\ref{f1}(b)). We adjust the barrier height of the quartic potential and assess the work extraction involved while transitioning the confinement shape.
The equilibrium probability distribution of particle position in the presence of such an external quartic potential can be derived as:
\begin{equation}\label{5}
    \begin{aligned}
      P_{eq}(x) &= \mathcal{N}_b  \exp \bigg [ \frac{\overline{a}}{2}x^2 - \frac{\overline{b}}{4} x^4 \bigg ], \\
      \mathcal{N}_b 
 &= \frac{e^{-\epsilon}}{\pi} \bigg ( \frac{2\overline{b}}{\epsilon} \bigg )^{\frac{1}{4}} \big [ I_{\frac{1}{4}} (\epsilon) +I_{-\frac{1}{4}} (\epsilon) \big ],
    \end{aligned}
\end{equation}
with $\epsilon = \Delta \overline{E}/2$. Here $I_{\nu}(z)$ is the modified Bessel function of the first kind of the form, $I_{\nu}(z) = \sum_{k=0}^{\infty} \frac{(z/2)^{2k+\nu}}{\Gamma [k+ \nu + 1] k!}$. Fig.~\ref{f2}(b) depicts the gradual changes in $P_{eq}(x)$ as the energy barrier is altered. 
In the limit of $\Delta \overline{E} \to 0$ $(a \to 0)$, $P_{eq}(x)$ shows an unimodal distribution reflecting the dominance of quartic term in potential, i.e. $V(x) \to \frac{b}{4} x^4$. With increasing the $\Delta \overline{E}$ (by an increase in $a$), $P_{eq}(x)$ shows bi-modality with two symmetric peaks and a distinct trough. Using Eq.~\ref{5}, the average extractable work $(-\langle \overline{W} \rangle)$ can be found influenced by the scaled barrier height $(\Delta \overline{E})$ as:
\begin{equation}\label{6}
    \begin{aligned}
-\langle \overline{W} \rangle =   & \frac{[4\epsilon - 1] I_{-\frac{1}{4}} \left( \epsilon \right)
-[4 \epsilon +1] I_{\frac{1}{4}}\left(\epsilon\right)}
{4  [ I_{\frac{1}{4}}\left( \epsilon \right ) +I_{-\frac{1}{4}}\left( \epsilon \right ) ]}
\\ &-\frac{ \epsilon [ I_{\frac{3}{4}}( \epsilon )+I_{\frac{5}{4}}\ ( \epsilon ) ]}{I_{\frac{1}{4}}( \epsilon )+I_{-\frac{1}{4}} (\epsilon)}.
 \end{aligned}
\end{equation}
As shown in Fig.~\ref{f4}, $-\langle \overline{W} \rangle$ decreases monotonically with increasing $\Delta \overline{E}$. 
In the limit of $\Delta \overline{E} \to 0$, the information-energy conversion is bounded by the $\frac{1}{4}k_BT$ as the monostable confinement is predominantly governed by quartic term. On raising $\Delta \overline{E}$, one can find that, beyond a specific scaled potential barrier, the system undergoes a refrigeration (cooling) process; we call this an inversion barrier ($\Delta \overline{E}_i$, scaled energy barrier). On solving the $-\langle \overline{W} \rangle = 0$ (using Eq.~\ref{6}), we determine the inversion barrier to be $\Delta \overline{E}_i \approx 0.24 $ regardless of the individual potential profile parameters or bath temperature (please see \cite{supple}). 

\begin{figure}
    \centering
        \centering
        \includegraphics[width=0.43\textwidth]{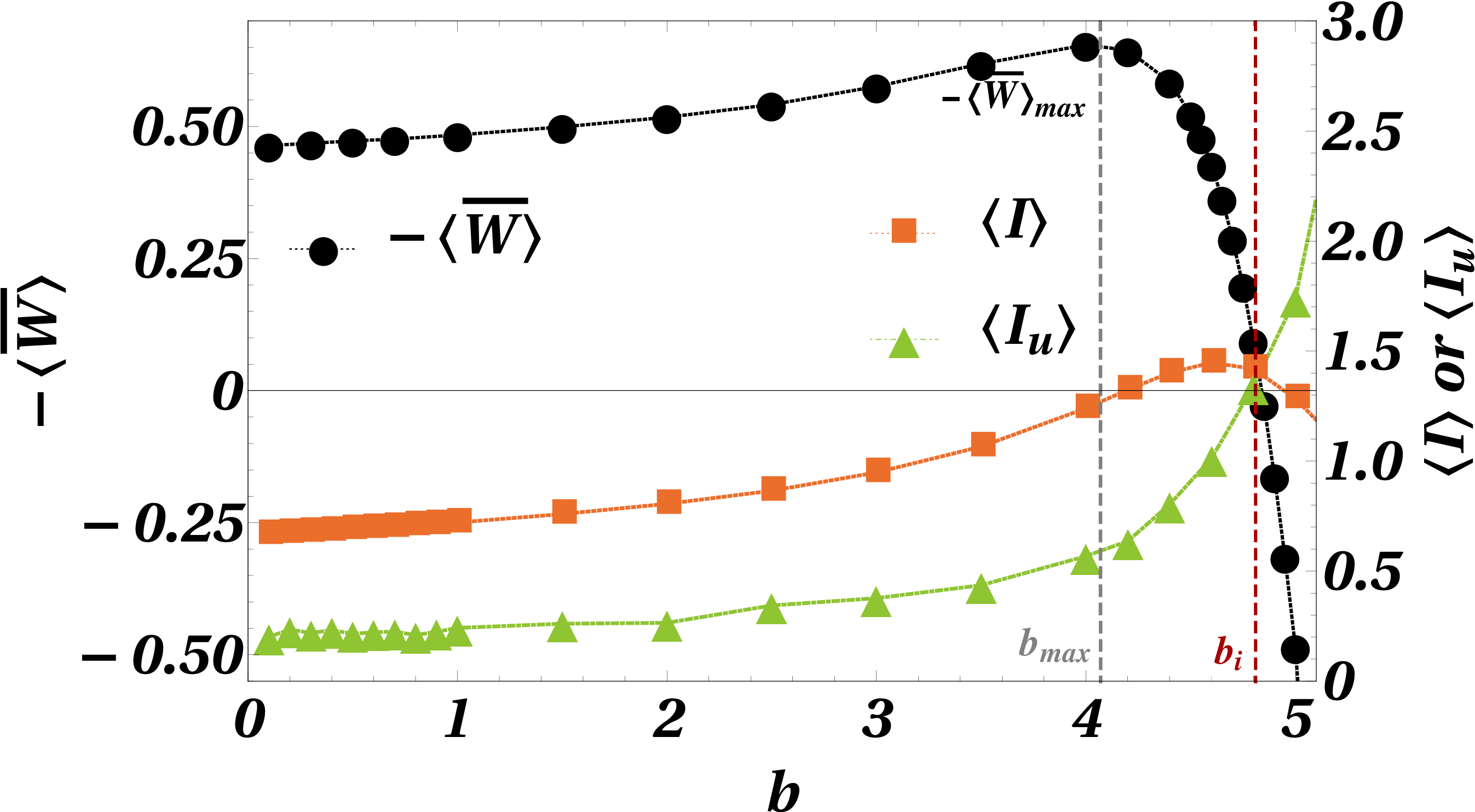} 
    \caption{The extracted work $(- \langle\overline{W }\rangle )$, information $(\langle I \rangle)$ and unavailable information $(\langle I_u \rangle)$ as function of $b$. For the parameter set chosen: $a = 4$, $c = 1$ and $D=1$. The points are obtained from numerical simulation (Eq.~\ref{1}).}
    \label{f5}
\end{figure}

The information gathered during the measurement step can be derived using Eq.~\ref{5} as:
\begin{equation}\label{7}
    \begin{aligned}
\langle I \rangle = -\ln & [ \mathcal{N}_b] + \frac{[4\epsilon - 1] I_{-\frac{1}{4}} \left( \epsilon \right)
-[4 \epsilon +1] I_{\frac{1}{4}}\left(\epsilon\right)}
{4  [ I_{\frac{1}{4}}\left( \epsilon \right ) +I_{-\frac{1}{4}}\left( \epsilon \right ) ]}
\\ &-\frac{ \epsilon [ I_{\frac{3}{4}}(\epsilon )+I_{\frac{5}{4}}\ (\epsilon ) ]}{I_{\frac{1}{4}}(\epsilon )+I_{-\frac{1}{4}} (\epsilon)} .
    \end{aligned}
\end{equation}
Similarly, unavailable information reads  as: $\langle I_u \rangle = - \ln [\mathcal{N}_b]$. As depicted in Fig.~\ref{f4}, the $\langle I \rangle$ and $\langle I_u \rangle$  increases with the potential barrier $(\Delta \overline{E})$. In the lower limit of energy barrier, $\Delta \overline{E} < \Delta \overline{E}_i$, the potential transitions from mono-stable to bistable as the barrier energy increases, leading to greater uncertainty in the particle's position. This will, in turn, increase $\langle I \rangle$ and $\langle I_u \rangle$, with $\langle I \rangle > \langle I_u \rangle$ resulting in positive available information and, thus, operating as an engine. For the double-well potential with  $\Delta \overline{E} = \Delta \overline{E}_i$, the information acquired during the measurement is equal to the information lost during relaxation phase $(\langle I \rangle = \langle I_u \rangle)$, leading to no work extraction. With further increase in barrier energy $\Delta \overline{E} > \Delta \overline{E}_i$, the potential wells get deeper. At stationary conditions, particles are more likely (less surprise) to be found rather near the stable basins than the potential centre (Fig.~\ref{f2}(b)). In this situation, information lost during the relaxation process exceeds the acquired information during the feedback, $(\langle I \rangle < \langle I_u \rangle)$, thereby acting as a refrigerator.\\ 
\noindent \textit{Profiling a triple well confinement:}
 Finally, we focus on the information processing of a BIE with multi-stable centrosymmetric confinement, keeping a stable well at the centre. A standard approach in developing such a potential profile is through a sextic potential of the form: $V(x) = \frac{ax^2}{2}-\frac{b x^4}{4}+\frac{c x^6}{6}$, where $a$, $b$ and $c$ are constant potential parameters. 
 As shown in Fig.~\ref{f1}(c), the potential shape can be modified from a mono-stable to a tri-stable energy landscape by continuously tuning the value of $b$ from a zero to non-zero one (for a given $a$ and $c$ value). Associated $P_{eq}(x)$ shows a variation from unimodal to trimodal steady-state distribution by changing the contribution of the quartic term, see Fig.~\ref{f2}(c). 
 The upper bound of the achievable average work, information and unavailable information can be calculated numerically using the formal definitions. As shown in Fig.~\ref{f5}, for a potential with a dominating quadratic term $(a>c)$, the output $-\langle W \rangle$ varies non-monotonically as a function of quartic term contribution. In the limit of $(b \to 0)$ and with $a \gg c$, the quadratic term dominates ($V(x) \sim \frac{a}{2}x^2$). Thus, one can obtain  $-\langle W \rangle \sim \frac{1}{2}k_BT$. The contribution of the $\frac{c x^6}{6}$ term determines the extent of decrease from $\frac{1}{2}k_BT$. With the introduction of the negative quartic term $(b>0)$, the shape of the potential arms acquires a partial concave shape (Fig.~\ref{f1}(c)). The influence of such concave shoulders on the potential can be realized in terms of the increased work output (for small but increasing values of $b$). For a further increase in $b$, the potential takes a proper triple well shape that alters (reduces) the overall information gain during the feedback and $-\langle W \rangle$ decreases consequently. This results in an interesting turnover of the work output for an optimal choice of $b$ when the potential shape consists of a mono-stable well at the centre and two centro-symmetric concave shoulders (looking similar to a shape with inflection points (please see \cite{supple})). For $b \to \text{ high }$, the influence of the quartic term intensifies, causing the cycle to function as a refrigerator.
\paragraph*{}The net acquired information $(\langle I \rangle)$ and unavailable information $(\langle I_u \rangle)$ can be calculated from the stationary state probability distribution and their variations as a function of quartic contribution $(b)$ has been shown in Fig.~\ref{f5} (right-hand side axis). The variations depict that $(\langle I \rangle)$ shows a non-monotonic turnover (likewise the $- \langle \overline{W} \rangle$) and $(\langle I_u \rangle)$  increases monotonically with increasing strength of the quartic contribution. The nonmonotonic change in $(\langle I \rangle)$ with increasing $b$ can be explained in a similar spirit as explained during the analysis of work output observed in the prescence of a monstable trapping. The presence of concave potential shoulders influences the increase in net surprise in particle position during the feedback process. While the potential has a proper triple well shape (high value of $b$), the information acquired during the feedback drops and the information loss during the relaxation process dominates over the former. Therefore, with the appropriate tuning of concavity and convexity of the potential profile, one can expect even more extractable work than monostable trapping. While analysing carefully, we find a few more interesting observations (as shown in \cite{supple}): the maximum output can be expected near the inflection point (though qualitative), the confinement with a dominating sextic term ($a \le c$) shows a monotonic decrease in work extraction with increasing $b$ and the maximum work extraction decreases with an increase in the quadratic term contribution $(c)$. The variation of the work performed under such different potential shaping has been described in the supporting information \cite{supple}. \\
\noindent \textit{Concluding remarks:}
We investigate the upper bound of the achievable work from a Brownian information engine, confined in external potentials of different shapes. We observe that harvested work varies with the nature of the potential landscape. For a BIE in single-well potential, $V(x)=a|x|^n,(n \ge 1)$, the upper bound of the achievable output work amounts to $\frac{1}{n} k_BT$. An increase in the concavity in the potential arms thus results in a more efficient information-energy exchange. To comprehend the impact of potential with unstable centers on the work, we explore the shift from a monostable to a bi-stable potential.  
As bi-stability is introduced, the upper bound of work decreases monotonically with increasing barrier height, ultimately leading the machine to transition into a refrigerator beyond a certain barrier height. To further understand the influence of multi-stability confinement on information processing, we examine the work extraction during the transition from a monostable to a tristable potential, featuring a stable central state and two symmetric stable basins.
During the transition, we observe a non-monotonic turnover in the upper bound output work. The early phase of transition results in increasing information harvesting, reaching peak output work for potential with concave shoulders. However, with well-defined triple-well trapping, the work decreases, and the machine eventually functions as a refrigerator. The enhanced work extraction can be accredited to a higher information gain due to the presence of a pseudo-concave potential surface.

\section*{acknowledgments}
 RR acknowledges IIT Tirupati for fellowship [DST/INSPIRE/03/2021/002138]. DM thanks SERB (Project No. ECR/2018/002830/CS), Department of Science and Technology, Government of India, for financial support and IIT Tirupati for the new faculty seed grant.
\section*{Data Availability}
The data that support the findings of this study are available within the article.
\section*{Conflict of interest}
The authors have no conflicts to disclose.
\bibliographystyle{achemso}
\bibliography{References}
\end{document}


\preprint{AIP/123-QED}

\title {Supplementary information: \\
Achievable Information-Energy Exchange in a Brownian Information Engine through Potential Profiling}

\author{Rafna Rafeek}
\author{Debasish Mondal}
  \email{debasish@iittp.ac.in}
\address{Department of Chemistry and Center for Molecular and Optical Sciences \& Technologies, Indian Institute of Technology Tirupati, Yerpedu 517619, Andhra Pradesh, India}

\date{\today}
\maketitle

\section{S1: Feedback protocols and definition of obsevables}
\begin{figure}[!htb]
    {\includegraphics[width=0.8\textwidth]{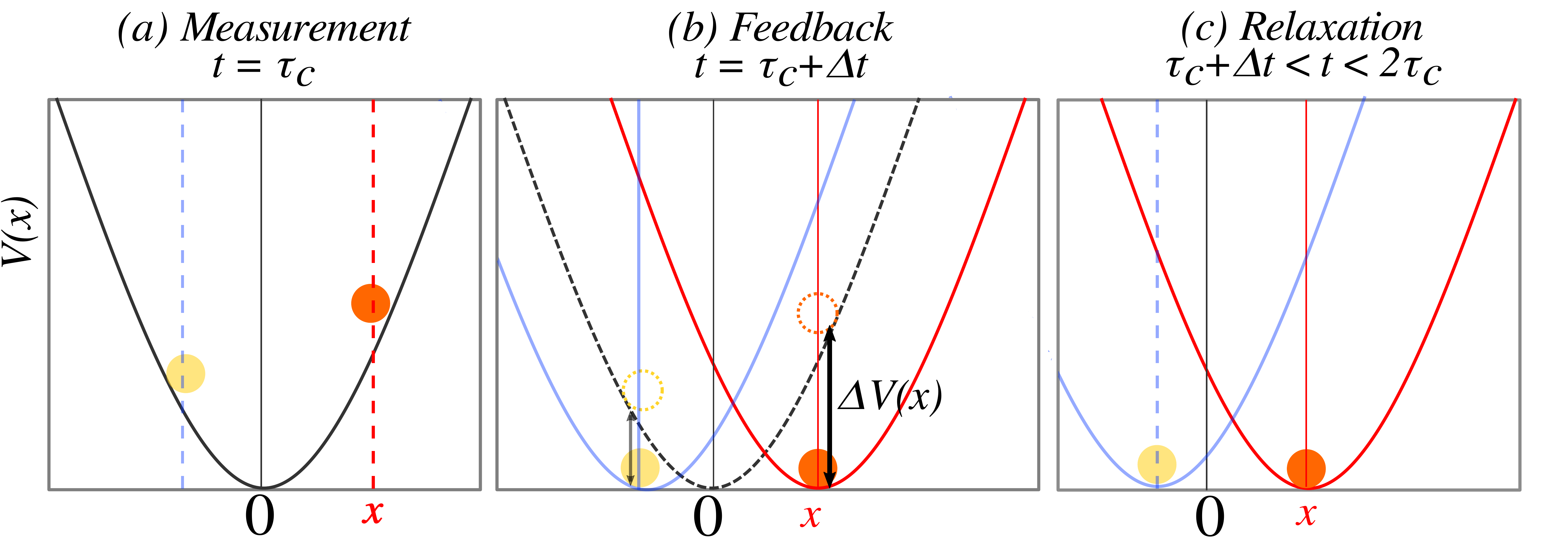}} 
     \caption{Schematic representation of engine cycle with Brownian particle trapped in monostable potential (forward-protocol). (a) In measurement step $( \lambda (\tau_c)=0 )$, particle position is measured accurately. (b) In the feedback step, according to the outcome, the potential center is instantly shifted $(\lambda(\tau_c + \Delta t)=x)$, and work is extracted from the potential difference. (c) In the relaxation step, the particle equilibrates in the shifted potential.}
    \label{sf1}
\end{figure}
The forward feedback protocol consists of three steps: measurement, feedback and relaxation. We have a one-dimensional Brownian particle confined in a potential centered at origin $(V(x - \lambda))$ \cite{Toyabe2010natphys, Ashida2014pre, ali2022geometric}. At the start of the engine cycle, the system is in thermal equilibrium with potential centered at origin $\lambda = 0$. In the measurement step, the particle position is estimated $(x(t)=x')$. Later in the next feedback step, the potential center is instantly changed from $\lambda = 0$ to $\lambda =x'$. Then, we allow the particle to relax back to thermal equilibrium with potential now centered at $\lambda = x'$, and the cycle repeats. We have demonstrated the forward protocol with particles confined in single well potential in Fig.~\ref{sf1}. \\

As the shift of the potential center is instantaneous, all the potential energy difference is converted into work, $-W(x') = \Delta V(x')= V(x')-V(0)$. The process is repeated and the average work obtained from the information engine is defined as:
\begin{equation}\label{s1}
        -\langle W \rangle = \int_{-\infty}^{\infty}dx' ( V(x')-V(0) ) P_{eq}(x') .
\end{equation}
Next, we calculate the net average information gained $(\langle I \rangle)$ during the feedback protocol. Given that we assume an error-free measurement, the net information acquired corresponds to the Shannon entropy of the particle in its initial equilibrium state. The average information gathered is given by:
\begin{equation}\label{s2}
        \langle I \rangle =- \int_{-\infty}^{\infty}dx'  P_{eq}(x') \ln P_{eq}(x'). 
\end{equation}
\begin{figure}
    {\includegraphics[width=0.8\textwidth]{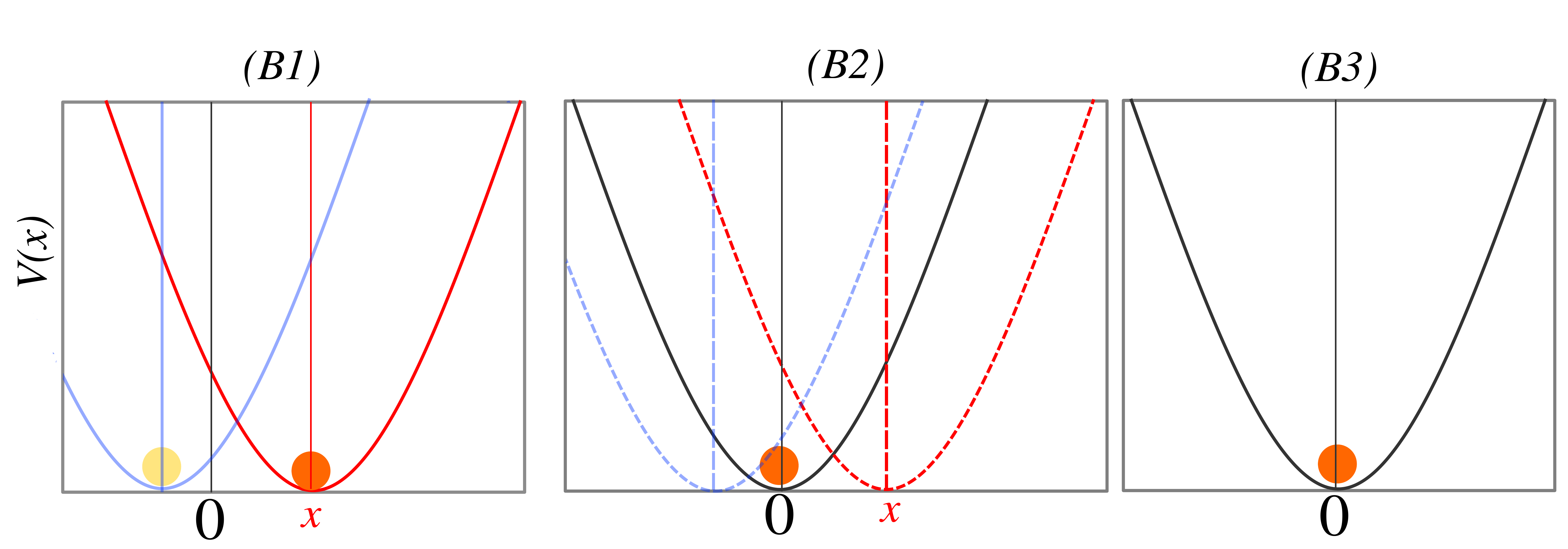}} 
     \caption{Schematic representation of backward feedback protocol with Brownian particle in a monostable confinement. In the initial step $(B1)$, the system is in equilibrium distribution of the final stage of the forward process. In the next steps $(B2,\;B3)$, the potential center is instantly shifted back to the initial potential center of the forward process $\lambda (t)= 0$. }
    \label{sf2}
\end{figure}
During the relaxation process, information about the particle position is lost. To quantify this loss of information, it is essential to examine the backward feedback protocol, as illustrated in Fig.~\ref{sf2}. In the reverse protocol: the system starts in equilibrium with the potential center at $\lambda = x'$. Then the potential center is instantaneously shifted to $\lambda = 0$. At this point, the probability of finding the particle at $x'$ is $P_{eq}(x)$. Consequently, the unavailable information is determined by averaging this probability as,
\begin{equation}\label{s3}
        \langle I_u \rangle =- \int_{-\infty}^{\infty}dx' \ln P_{eq}(0) P_{eq}(x').
\end{equation}
\section{S2: Parameters and work variations related to Potential Profiling}

\begin{figure}[!htb]
    {\includegraphics[width=0.45\textwidth]{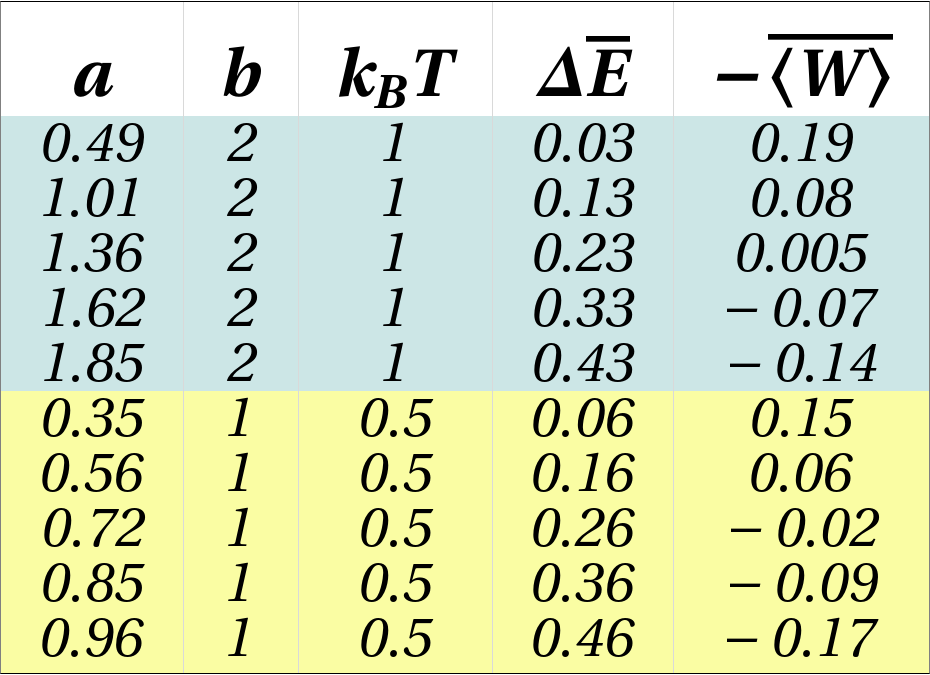}} 
     \caption{Summary of  extracted work $(- \langle\overline{ W} \rangle)$ obtained at different values of $a$, $b$ and $k_B T$. Here $\Delta \overline{E}$ is the scaled barrier height of double-well potential with the given parameter set. With other parameter set chosen: $c=1$, $\gamma = 1$ and $D=1$.}
    \label{sf3}
\end{figure}

\noindent \paragraph{Double-well confinement study:}
In the Fig. 3 of the main text, we find a monotonic decrease in scaled work $(- \langle\overline{ W} \rangle)$ with increasing scaled energy barrier $(\langle\overline{E}\rangle)$. $\Delta \overline{E}$ corresponds to different combinations of double-well potential parameter $(a,b)$ and bath temperature $(T)$. Here in Fig.~\ref{sf3}, we present a tabular summary of scaled average work $(- \langle\overline{ W} \rangle)$ obtained for different values of scaled barrier height $(\Delta \overline{E})$.

\paragraph{Triple-well Confinement study:}
\begin{figure}
    \centering     \includegraphics[width=0.45\textwidth]{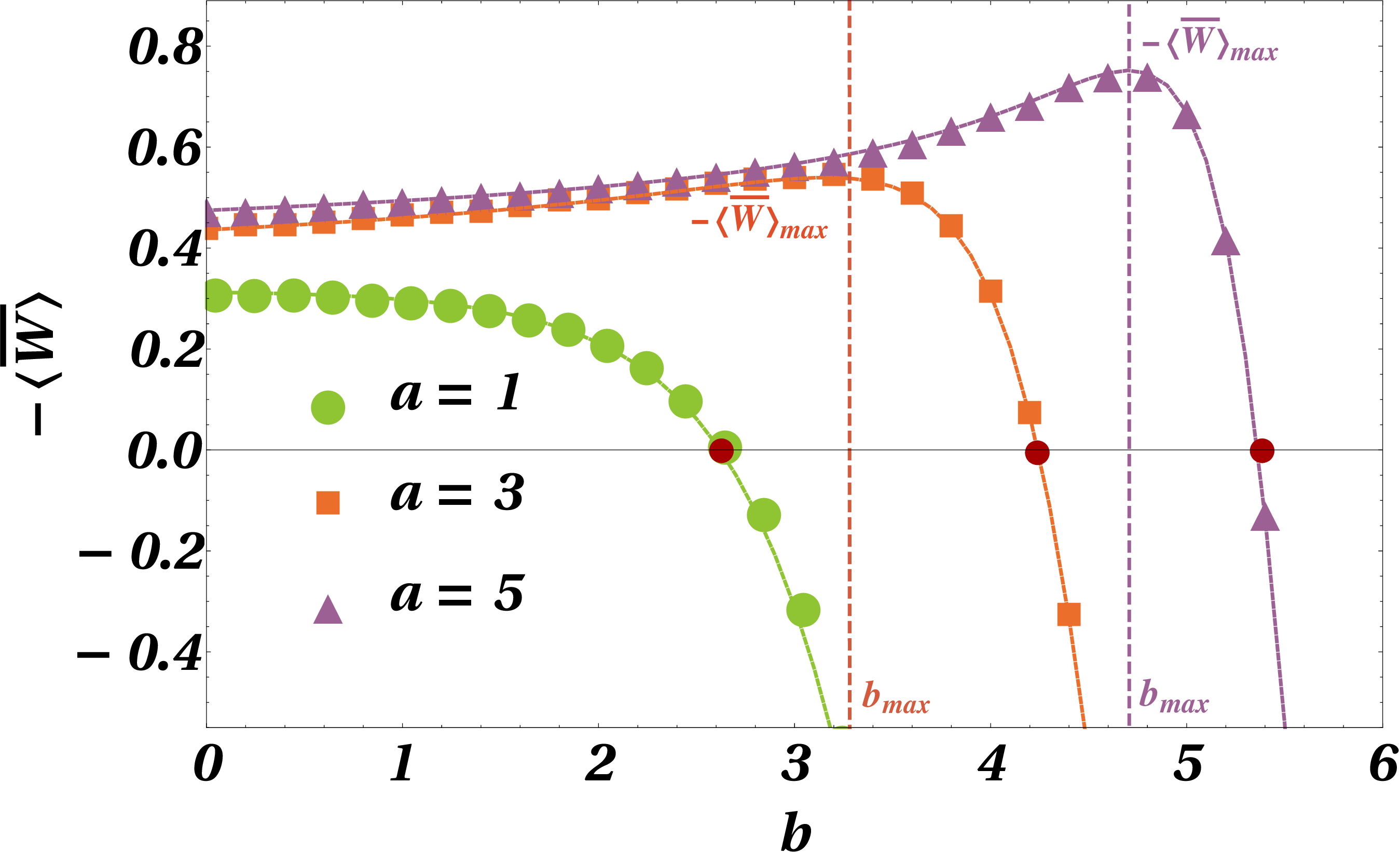} 
    \caption{The variation of extracted work $(- \langle\overline{ W} \rangle)$ as a function of $b$  for different values of $a$. With other parameter set chosen: $c=1$, $\gamma = 1$ and $D=1$. The points are obtained from numerical simulation .}
    \label{sf4}
\end{figure}
The achievable work extraction can be numerically calculated using the definition of work extraction. The Fig.~\ref{sf4}, shows the variation in the amount of work output as a function of quartic term contribution $(b)$ for different sets of quadratic influence $(a)$. The potential confinement with a dominating sextic term $(a \leq c)$ shows a monotonic decrease in work extraction with increasing $b$. However, the potential profile dominated by quadratic term $(a>c)$ shows a non-monotonic variation of work output. Therefore we define $- \langle \overline{W} \rangle_{max}$ as maximum work obtained at a specific value of $b_{max}$, given a constant set of $a$ and $c$. The maximum work extraction $(- \langle \overline{W} \rangle_{max})$ increases with quadratic term contribution $(a)$. Potential profiles with dominant sextic $(c \ge a)$ influence to a monotonic decrease in work extraction with increasing quartic term contribution.

\begin{figure}[!htb]
    {\includegraphics[width=0.45\textwidth]{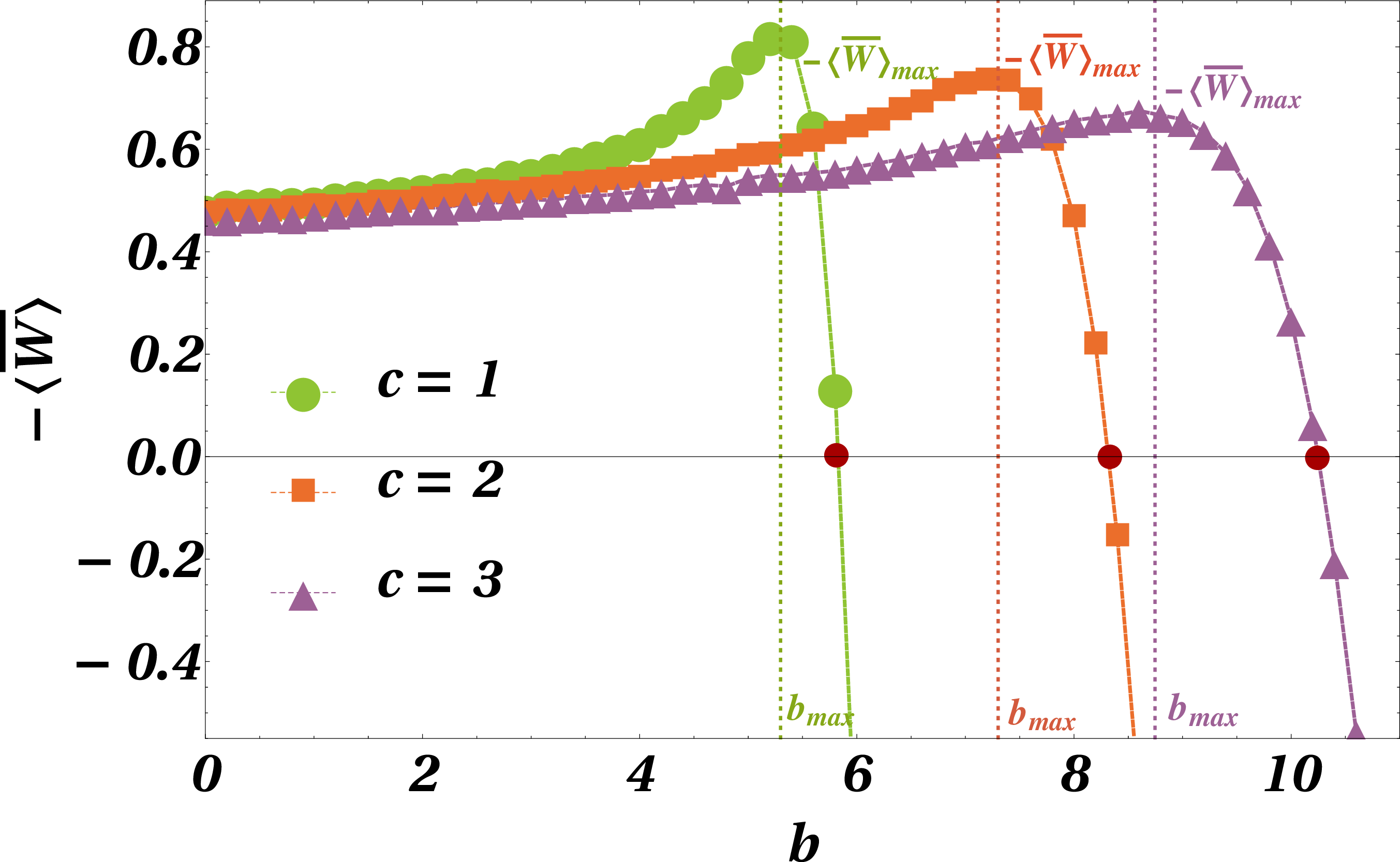}} 
     \caption{The extracted work $(- \langle\overline{ W} \rangle)$ as function of $b$  for different values of $a = 4$. With other parameter set chosen: $\gamma = 1$ and $D=1$. The points are obtained from numerical simulation.}
    \label{sf5}
\end{figure}

The Fig.~\ref{sf5}, shows the variation in the amount of work output as a function of quartic term contribution $(b)$ for different sets of sextic influence $(c)$. The potential confinement with a dominating quadratic term $(a > c)$ shows a non-monotonic variation in work extraction with increasing quartic term contribution. The maximum work extraction $(- \langle \overline{W} \rangle_{max})$ decreases with an increase in the quadratic term contribution $(c)$. \\

\begin{figure}[!htb]
    {\includegraphics[width=0.45\textwidth]{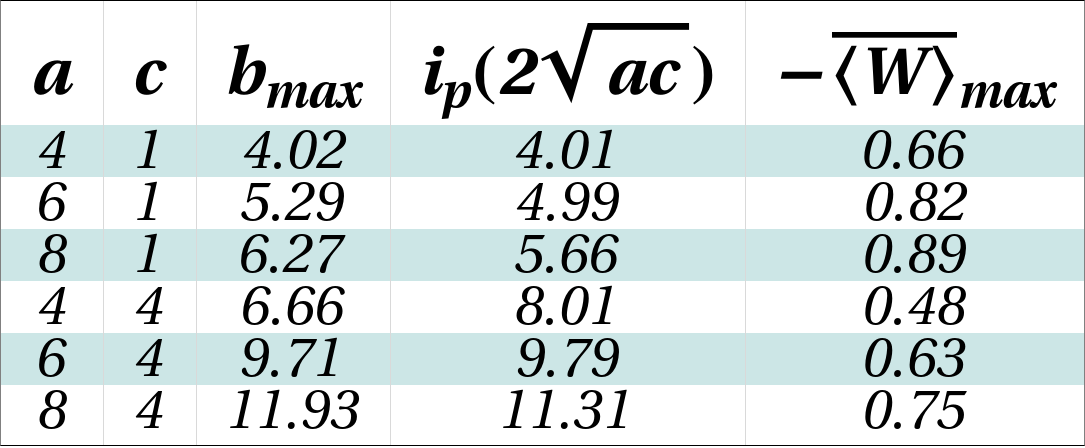}} 
     \caption{Summary of maximum extracted work $(- \langle\overline{ W} \rangle_{max})$ obtained at $b_{max}$ for different values of $a$ and $c$. Here $i_p$ is the inflection point of the triple-well potential with the given parameter set. With other parameter set chosen: $\gamma = 1$ and $D=1$. The data obtained from numerical simulation.}
    \label{sf6}
\end{figure}

Fig.~\ref{sf6}, is a tabular summary of scaled maximum work $(- \langle \overline{ W} \rangle_{max})$ obtained and corresponding $b_{max}$ for potential profiles with different sets of $a$ and $c$. Here inflection point $i_p =2\sqrt{ac}$ denotes the quartic contribution $(b)$ for a given set $a$ and $c$ resulting in a metastable potential profile. From the table summary, it is evident that for potential landscape with dominating quadratic term $a > c$ maximum work extraction is obtained for potential confinement with $b_{max} \approx i_{p}$.

\bibliographystyle{achemso}
\bibliography{References}